# Properties of the non-Gaussian fixed point in 4D compact U(1) lattice gauge theory

J. Cox[a], W. Franzki[a], J. Jersák[a], C. B. Lang,[b], T. Neuhaus[*c] and P. Stephenson[d]

[a]Institut für Theoretische Physik E, RWTH Aachen, Germany

[b]Institut für Theoretische Physik, Karl-Franzens-Universität Graz, Austria

[c]FB8 Physik, BUGH Wuppertal, Germany

[d]IFH Zeuthen, DESY, Germany

We examine selected properties of the gauge-ball spectrum and fermionic variables in the vicinity of the recently discussed non-Gaussian fixed point of 4D compact U(1) lattice gauge theory within the quenched approximation. Approaching the critical point from within the confinement phase, our data support scaling of $T1^{+-}$ gauge-ball states in units of the string tension square root. The analysis of the chiral condensate within the framework of a scaling form for the equation of state suggests non mean-field values for the magnetic exponents $\delta$ and $\beta_{exp}$.

## 1. Introduction

The strongly coupled $4D$ compact U(1) lattice gauge theory has a non-Gaussian fixed point, whose critical behavior is governed by a correlation length critical exponent $\nu = 0.365(8)$ [1–3]. Introducing the double-charge coupling $\gamma$ Wilsons action is extended to

$$S = -\sum_P [\beta \, cos(\Theta_P) + \gamma \, cos(2\Theta_P)]. \qquad (1)$$

For values of the double charge coupling $\gamma \leq 0$ the critical singularity at the confinement-Coulomb phase transition is governed by the non-Gaussian fixed point. This opens the very interesting possibility for the construction of an interacting $4D$ field theory at finite values of the renormalized charge. The physical content of the theory at the confinement-Coulomb phase transition depends on the phase from which the critical line is approached. In the confinement phase a confining theory with monopole condensate is expected. The physical spectrum consists of various gauge-balls. In the Coulomb phase massless photons and massive magnetic monopoles should be present. With staggered dynamic fermion degrees of freedom included the lattice theory is known to have both the confinement and the Coulomb phases. In the confinement phase the chiral symmetry is broken quite analogous to QCD. If also at finite fermion mass-values $m_0$ -or even at $m_0 = 0$- the new non-Gaussian fixed point is encountered, then the corresponding continuum theory might -or might not- share many aspects of QCD. It may also contribute to the problem of the Landau pole and triviality of QED. This considerations strongly motivate a study of compact QED with staggered Kogut Susskind fermions included. Here we consider the case of quenched fermion degrees of freedom, which may be viewed as the first step towards our ambitious goal of constructing the continuum limit of full QED.

The considerations leading to the new non-Gaussian fixed point rely on a regularization of the pure gauge theory on homogeneous spherical lattices, lattices with a sphere like topology. This way spurious two-state signals can be eliminated at criticality in the numerical simulation of the path integral. For the spectrum calculations, and also for the implementation of 4-flavor staggered quenched fermions, we study the theory on torodial lattices, which are much simpler than spherical lattices. We argue that effects related to the topology of the lattice can be neglected, if the theory is studied in a finite distance from the critical point. Our simulations show, that correlation length values of about few lattice spacings can be simulated in a parameter region of the theory, where the effects of topology are negligible i.e., in a region without spurious metastability signals.

[*]Speaker at the conference.

In this report we limit our presentation to a study of the confinement phase at a value $\gamma = -0.2$ for the double charge coupling. All data are calculated on a large $16^3 \times 32$ lattice. For a detailed study of the theories spectrum [4] and its fermionic properties [5] we refer to two forthcoming publications.

## 2. Results

We implemented measurements of the gaugeball spectrum by considering the representations of the cubic group [6]. Smearing technology for the purpose of improving ground state overlaps was applied [7]. Two values of the spacial momentum $k_x = 0$ and $k_x = 1$ have been considered. In figure 1 we display masses of the $T1^{+-}$-state as a function of $\beta$ (triangles) in comparison with the string tension square root (circles).

Both quantities exhibit a scaling approach within a scaling window $\propto (\beta_c - \beta)^{0.365}$, consistent with the correlation length divergence expected at the non-Gaussian fixed point, the solid curves in figure 1. They are obtained by a fit with fixed value for the correlation length divergence exponent to the solid symbols.

The estimated value of the $T1^{+-}$-mass in units of the string tension square root is about 3.

In the confinement phase we study the chiral condensate $\Sigma = <\bar{\chi}\chi>$ in the framework of a scaling form for the equation of state

$$am_0 = \Sigma^\delta F\left(\frac{t}{\Sigma^{\frac{1}{\beta_{exp}}}}\right), \qquad (2)$$

where $m_0$ denotes the bare fermion mass. Such a scaling form is based on the analogy of chiral symmetry breaking to magnetic materials and is expected to hold in the vicinity of the critical point, $t$ denoting the distance from criticality $t = \beta_c - \beta$. Mean field values for the "magnetic" exponents are $\beta_{exp} = \frac{1}{2}$ and $\delta = 3$. Finite size effects to the considered quantity are very small and we rely here on data from a $16^3 \times 32$ lattice. We consider up to six bare mass-values $m_0 = 0.005, 0.010, 0.020, 0.035, 0.050$ and $0.100$.

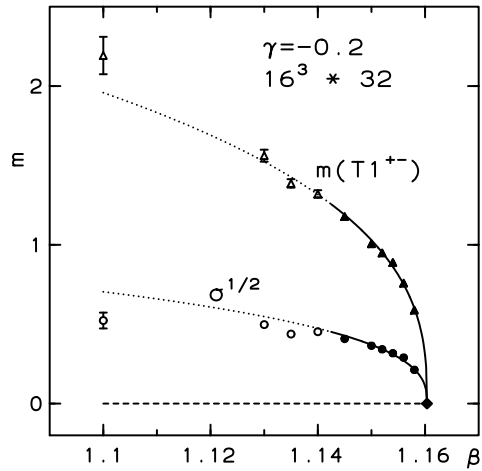

Figure 1. Masses of $T1^{+-}$ state in comparison with $\sigma^{\frac{1}{2}}$.

In figure 2 we examine a possible mean-field value for the exponent $\delta$ by plotting the square chiral condensate $\Sigma^2$ as a function of the quantity $am_0/\Sigma$ for several values of $\beta$ and $m_0$. Deviations from a straight line behavior at fixed $\beta$ are clearly visible in the vicinity of the critical point. Thus a mean-field exponent-value is ruled out. Choosing the set of parameters $\beta_{exp} = \frac{1}{3}$, $\delta = 1.9$ and $\beta_c = 1.1609$ all data collapse on a single scaling curve, the function F(x), if the quantity $am_0/\Sigma^\delta$ is plotted as a function of $-x = -t/\Sigma^{\frac{1}{\beta_{exp}}}$. The situation is depicted in figure 3. Our findings strongly suggest a non-linear form for the scaling function $F(x)$. The quoted exponent values differ considerable from mean-field values. Currently we regard the quoted values only as approximative. The final analysis for the exponent-values

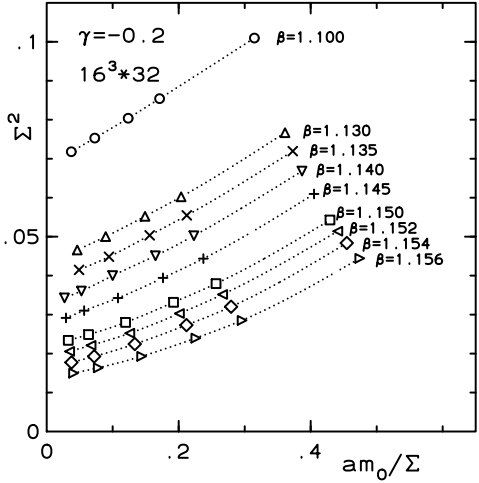

Figure 2. Scaling plot as described in the text at a mean-field value of $\delta$.

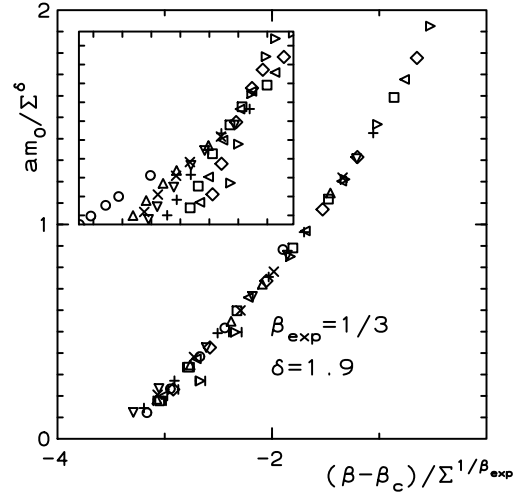

Figure 3. The scaling function $F(x)$ for non mean-field values of $\beta_{exp}$ and $\delta$. The inlay of the figure depicts the situation for mean-field values $\beta_{exp}$ and $\delta$. No scaling form is observed.

will be presented elsewhere.

## 3. Summary

A study of the gauge-ball spectrum and quenched fermionic observables on torodial lattices appears to be feasable at the new non-Gaussian fixed point of 4D compact U(1) lattice gauge theory. As the critical point is approached within the confinement phase, $T1^{+-}$ states scale in units of the string tension square root. The corresponding correlation length divergence exponent is consistent with the non-Gaussian value at the fixed point. Chiral symmetry breaking can be described by a non mean-field and non-linear scaling form for the equation of state. A full account of the studies initiated in this paper will be presented soon [4,5]. We remark, that signals for a critical behavior in U(1) lattice gauge theory have also been reported by another group [8].

## REFERENCES

1. J. Jersák, C. B. Lang and T. Neuhaus, *Non-Gaussian fixed point in four-dimensional pure compact U(1) gauge theory on the lattice*, hep-lat/9606010, UNIGRAZ-UTP-29-05-96, to appear in Phys. Rev. Lett..
2. J. Jersák, C. B. Lang and T. Neuhaus, *Four-dimensional pure compact U(1) gauge theory on a spherical lattice*, hep-lat/9606013, UNIGRAZ-UTP-11-06-96, to appear in Phys. Rev. D.
3. C. B. Lang and P. Petreczky, *U(1) gauge theory with Villain action on spherical lattices*, hep-lat/9607038, UNIGRAZ-UTP-18-07-96.
4. J. Cox, W. Franzki, J. Jersák, C. B. Lang, T. Neuhaus and P. Stephenson, in preparation.
5. J. Cox, W. Franzki, J. Jersák, C. B. Lang and T. Neuhaus, in preparation.
6. B. Berg and A. Billoire, Nucl. Phys. **B221** (1983) 109.
7. C. Michael and M. Teper, Phys. Lett. **206B** (1988) 299; Nucl. Phys. **B314** (1989) 347.
8. W. Kerler, C. Rebbi and A. Weber, *Critical behavior and monopole density in U(1) lattice gauge theory*, hep-lat/9607009, BU-HEP-96-18.